\documentclass[%
 reprint,
nofootinbib,
 amsmath,amssymb,
 aps,
]{revtex4-1}

\usepackage{graphicx}% Include figure files
\usepackage{dcolumn}% Align table columns on decimal point
\usepackage{bm}% bold math
\begin{document}

\begin{flushright}
 LA-UR-19-22136 v2
\\
\end{flushright}
%\preprint{APS/123-QED}

\title{Vacuum Texture: A New Interpretation of Quantum Mechanics and a New Loophole for Bell's Inequality Measurements that preserves Local Realism and Causality}
\author{Yoko Suzuki}
 \email{ysuzuki@newmexicoconsortium.org}
\affiliation{
 New Mexico Consortium, Los Alamos, NM 87544, USA
}
\author{Kevin M Mertes}
 \email{kmmertes@lanl.gov}
\affiliation{
Los Alamos National Laboratory, Los Alamos, NM 87545, USA
}
\date{\today}

\begin{abstract}
We introduce a new interpretation of quantum mechanics by examining the Einstein, Podolsky and Rosen (EPR) paradox and Bell's inequality experiments under the assumption that the vacuum fluctuation has a locally varying texture (a local variable) for energy levels below the Heisenberg time-energy uncertainty relation. In this article, selected results from the most reliable Bell's inequality experiments will be quantitatively analyzed to show that our interpretation of quantum mechanics creates a new loophole in Bell's inequality, and that the past experimental findings do not contradict our new interpretation. Under the vacuum texture interpretation of quantum mechanics in a Bell's inequality experiment, the states of the pair of particles created at the source (e.g. during parametric down conversion) is influenced by an inhomogeneous vacuum texture sent with the speed of light from the measurement apparatus.  We will also show that the resulting pair of particles are \textbf{not} entangled and that the theory of vacuum texture preserves local realism with complete causality. This article will also suggest an experiment to definitively confirm the existence of vacuum texture.
\begin{description}
\item[PACS numbers]
03.65.-w, 03.65.Ta, 03.65.Ud
\end{description}
\end{abstract}

\pacs{Valid PACS appear here}
\maketitle

\section{\label{sec:INTRODUCTION}INTRODUCTION}

It has been a century since quantum mechanics has taken the central role in physics. Ever since Einstein, Podolsky and Rosen's (EPR) \cite{EPR} question was unexpectedly answered through experimental measurements \cite{Clauser, AspectJuly, AspectDec} of Bell's inequality \cite{Bell}, discussion of the interpretation of quantum mechanics has waned. In the past few decades, Bell's inequality measurements have been performed with high precision detection and control.  These measurements have repeatedly confirmed the result and have seemingly precluded many possible loopholes. After a whole century has passed, the vast majority of the scientific community seems to completely accept that quantum mechanics is a complete theory in the sense that there are no hidden variables. Perhaps, the findings from the Bell's inequality measurements are the main reasons why hidden variable interpretations are no longer considered as a viable means. A hidden variable theory is still the most likely way to model physical nature with local realism and causality. Here, we would like to revisit the EPR question \textit{one more time} to see what Bell's inequality and its past measurements are really telling us and whether we really have to give up on local realism and causality.

This article will shed light on an assumption that has been carelessly granted; the vacuum was originally assumed to be empty, calm and non-interacting. There have been some hints, such as zero-point energy, vacuum fluctuation, virtual particles through dark energy, that the vacuum should rather be full, stormy and interacting. Ultimately, quantum field theory interprets the vacuum as the completely filled ground states with random and uncertain fluctuations. Our model of the vacuum assumes it is neither empty nor full, but rather possessing a dynamic and deterministic texture (local variable). The details of the theory underlying the interpretation will be shown elsewhere (ref. \cite{Yoko1}). 

Here, we propose that, instead of vacuum fluctuations being completely stochastic, random and uncertain, they are locally deterministic and the fluctuations posses an inhomogeneous texture. The texture is created by the presence of matter (such as an experimental apparatus and the sample itself) and the texture travels with the speed of light (or very close to it). These low level fluctuations are hard to detect since they don't possess energy larger than what the Heisenberg time-energy uncertainty principle allows. However, vacuum texture can drastically break symmetry during particle creation. 

In this interpretation of quantum mechanics, the probability distribution is a consequence of an inhomogeneous vacuum texture and the particle is created in one state at the source and travels in one trajectory. The detection of the particle does not ``collapse'' its quantum state as suggested by the Copenhagen interpretation; rather, the placement of the detector has already broken the symmetry of the vacuum texture and thus changed the probability of the creation of a particle (and it's state) along various trajectories at the source. When the experimental apparatus configuration changes, the vacuum texture is adjusted with (near) luminous speed. In other words, in this interpretation, quantum information is stored in the vacuum texture. The quantum behavior is the consequence of measuring the vacuum texture directly or particles interacting with vacuum texture. A rigorous formulation of the theory of vacuum texture is beyond the scope of this article but will be discussed in ref. \cite{Yoko1}.

In this article, we will focus on the new vacuum texture loophole rather than the mechanism behind the interpretation. We will suggest an experiment similar to the one performed by Alain Aspect in 1982 \cite{AspectDec}; however, additional data analysis is needed in order to test the existence of vacuum texture. If the existence of vacuum texture is verified by an experiment, it will bring a new answer to the EPR question and will inevitably allow local variable theories with local realism and causality to be placed back on the table.

\section{\label{sec:Bell}Bell's Inequality with vacuum texture}

\begin{figure}[h]
\centering
\includegraphics[width=0.45\textwidth]{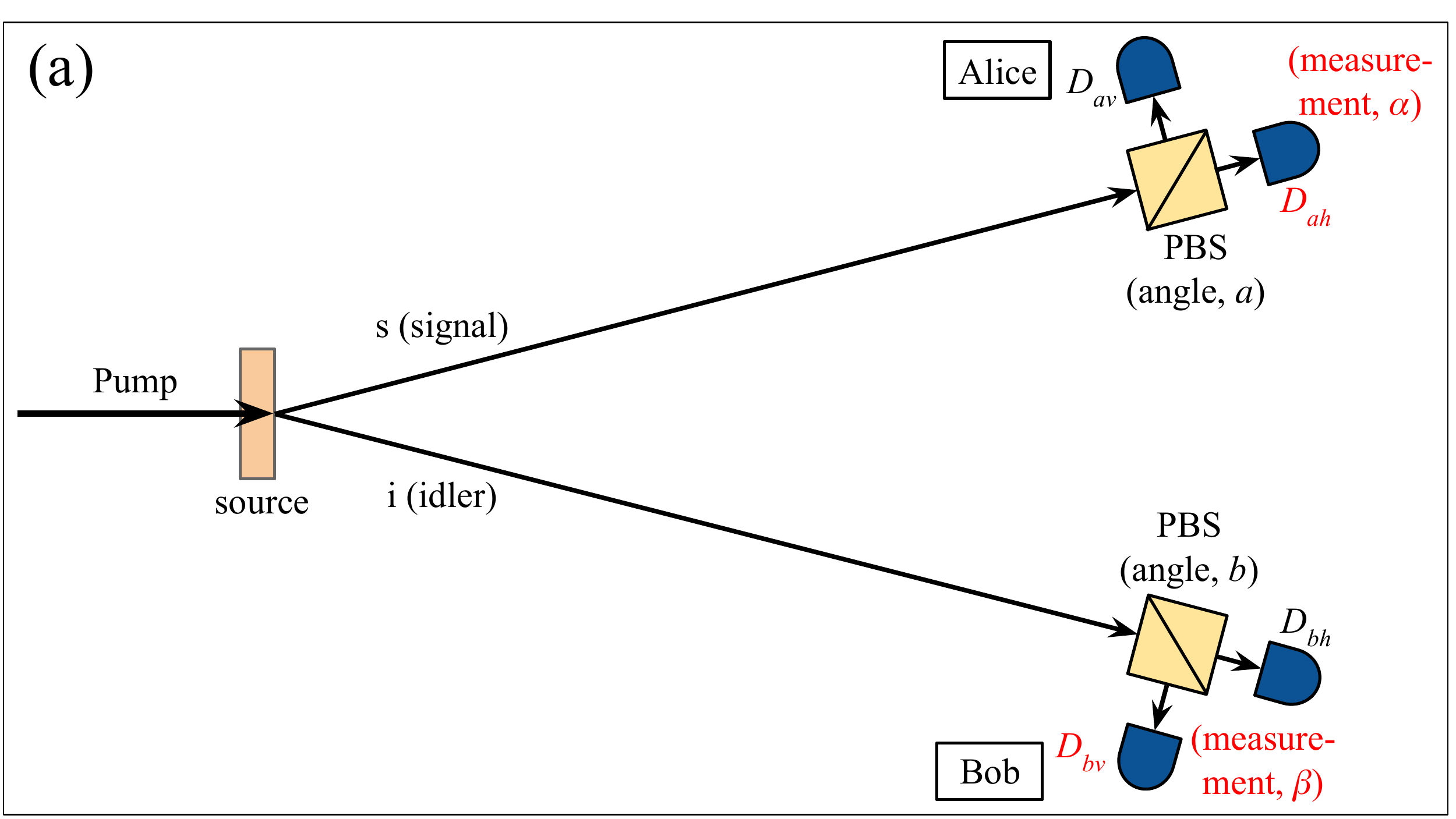}
\includegraphics[width=0.45\textwidth]{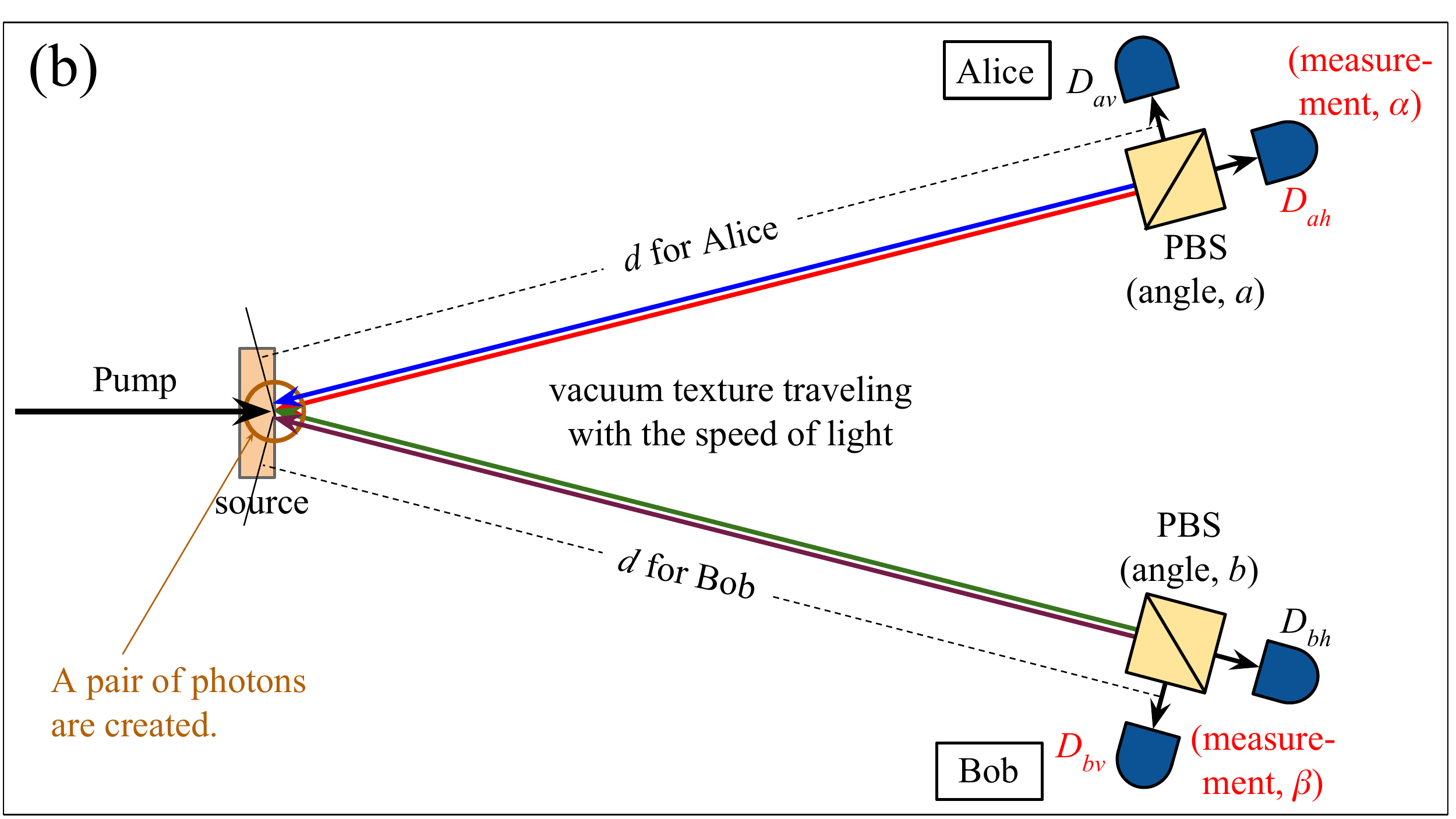}
\caption{(a) The typical Bell's inequality measurement setting for a pair of photons with a spontaneous-parametric-down-conversion photon source and polarization beam splitters (PBSs). (b) Under the assumption of vacuum texture, placing an object in a light-like cone that includes the source of the creation of a pair of photons effects the measurement and rotational symmetry at the source is broken. In term of conventional quantum optic, it could be interpreted as the squeezed vacuum state which no longer has symmetric distribution of uncertainty due to the measurement apparatus between the PBS and the source.}
\label{fig:settingAB}
\end{figure}

\begin{figure}[h]
\centering
\includegraphics[width=0.45\textwidth]{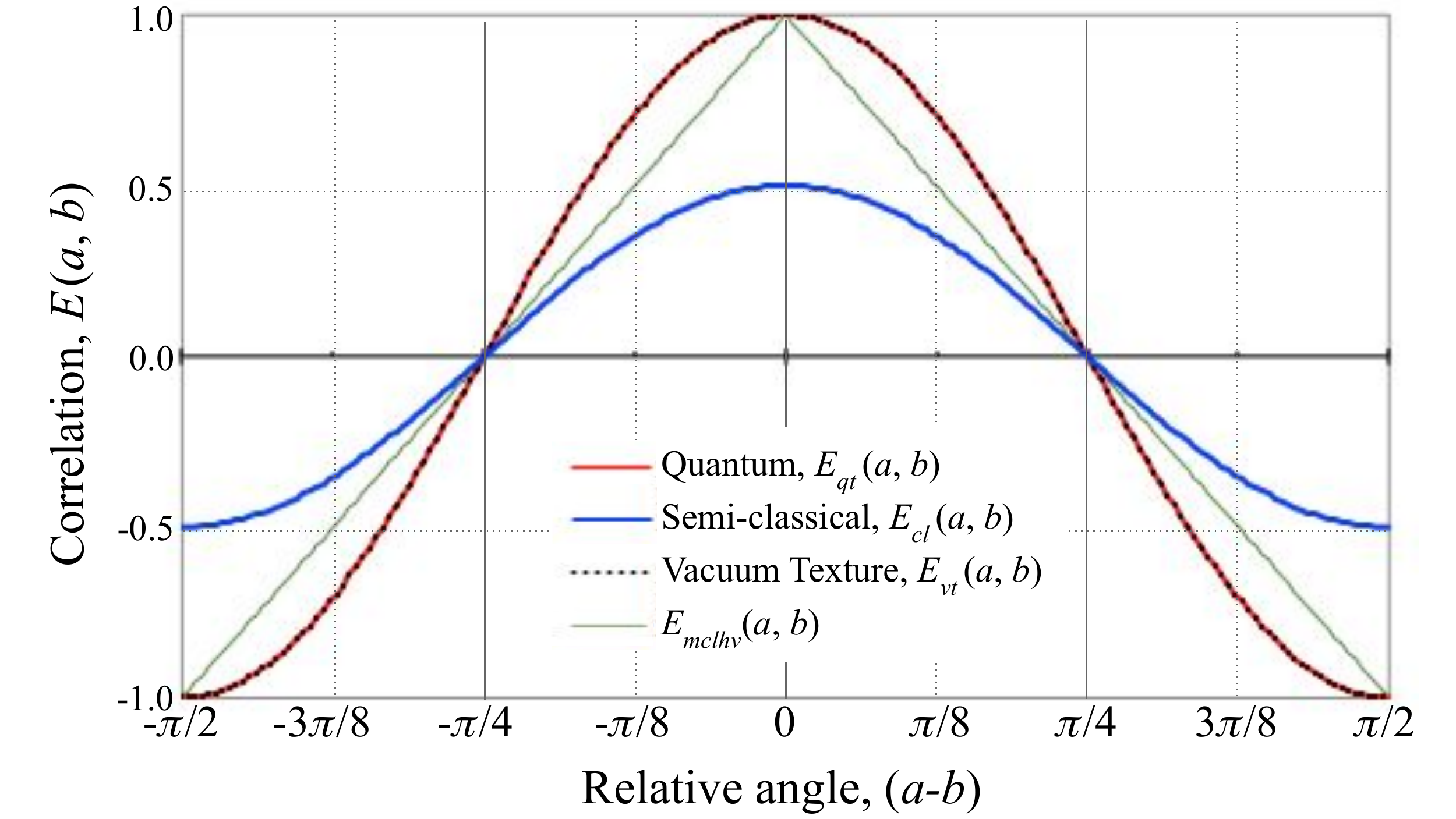}
\caption{$E_{qm}(a,b)$ is the quantum mechanical correlation function with Alice and Bob's measurements with settings $a$ and $b$ (red curve). $E_{sc}(a,b)$ is for the semi-classical solution using Eq. \eqref{eqn:lhv} (blue curve). $E_{vt}(a,b)$ is the semi-classical solution with vacuum texture using a local hidden variable in Eq. \eqref{eqn:Pvt} (black dotted curve, which is identical to the red curve). $E_{mclhv}(a,b)$ is maximum classical correlation using \eqref{eqn:lhv} when replacing Eq. \eqref{eqn:semic} with $\sum_{\alpha =\pm 1}\alpha P(\alpha|a,\lambda )=sgn(\cos(2(a-\lambda)))$ and 
$\sum_{\beta =\pm 1}\beta P(\beta|b,\lambda )=sgn(\cos(2(b-\lambda)))$. }
\label{fig:Angle}
\end{figure}

In a typical Bell's experiment, a source sends two physical systems (photon's in this case) to distant observers, Alice and Bob (Fig. \ref {fig:settingAB} (a)). The measurement setting chosen by Alice is $a$ and by Bob is $b$ while the outcome received by Alice is $\alpha$ and by Bob is $\beta$. For the case of photon polarization, $a$ and $b$ correspond to polarization beam splitter (PBS) angles. $\alpha$, $\beta=+1$ for detecting a photon and $-1$ for not detecting a photon. The experimentally measured probability distribution of the correlation between Alice and Bob produces results that violate Bell's inequality and are consistent with quantum entangled states. The correlation function is expressed as:
\begin{equation}
E(a,b)=\sum_{\alpha,\beta=\pm 1}\alpha \beta P(\alpha \beta|ab)
\label{eqn:Eab}
\end{equation}
where $P(\alpha \beta|ab)$ is the probability of the measurement result for $\alpha$, $\beta=+1$ or $-1$ given that Alice's and Bob's polarization measurement axis is at $a$ and $b$ degrees, respectively. 
With standard quantum mechanical analysis, Eq. \ref{eqn:Eab} becomes: 
\begin{equation}
E_{qm}(a,b)=\cos(2(a-b)),
\label{eqn:eqt}
\end{equation}
and is shown in Fig. \ref{fig:Angle} by the red curve.

For local hidden variable theories, the probability is expressed as:
\begin{equation}
P(\alpha \beta|ab)=\int_{\Lambda}d\lambda q(\lambda )P(\alpha|a,\lambda )P(\beta|b,\lambda),
\label{eqn:lhv}
\end{equation}
where, $q(\lambda)d\lambda $ is the distribution of the hidden variable, $\lambda$, and $\Lambda$ represents the manifold of allowed values for $\lambda$. For photon polarization, $\lambda$ is an angle and also $\int^{\pi/2}_{-\pi/2}d\lambda q(\lambda )=1$ where $q(\lambda)=q(\lambda \pm \pi)$.

In semi-classical analysis, assuming rotational symmetry during the creation of a pair of photons at the source, $q(\lambda )=1/\pi$, and with Malus's law,
\begin{equation}
\begin{split}
\sum_{\alpha =\pm 1} \alpha P(\alpha|a,\lambda )d\lambda=\cos(2(a-\lambda))d\lambda,\\
\sum_{\beta=\pm 1} \beta P(\beta|b,\lambda )d\lambda=\cos(2(b-\lambda))d\lambda.
\end{split}
\label{eqn:semic}
\end{equation}
Using Eq. \eqref{eqn:Eab}, this yields $E_{sc}(a,b)=(\cos{2(a-b)})/2$, which is depicted by the blue curve in Fig. \ref{fig:Angle}.

Bell's inequality is derived from Eq. \eqref{eqn:lhv} assuming that $q(\lambda )$ is independent of $a$ and $b$: $q(\lambda |ab)=q(\lambda )$. Within vacuum texture theory, this is no longer true because the value of the hidden variable depends on $a$ and $b$.  Thus, the probability of the measurement must be expressed as:
\begin{equation}
P_{vt}(\alpha \beta|ab)=\int^{\pi/2}_{-\pi/2}d\lambda q(\lambda |ab)P(\alpha|a,\lambda )P(\beta|b,\lambda).
\label{eqn:Pvt}
\end{equation}
Bell's inequality does not hold for Eq. \eqref{eqn:Pvt} because it relies on Eq. \eqref{eqn:lhv}. \textit{Eq. \eqref{eqn:Pvt} can be true even if Bell's inequality is violated; this is the new loop-hole that the theory of vacuum texture reveals.}

\begin{figure}[h]
\centering
\includegraphics[trim={2cm 1cm 2cm 0cm},width=0.35\textwidth]{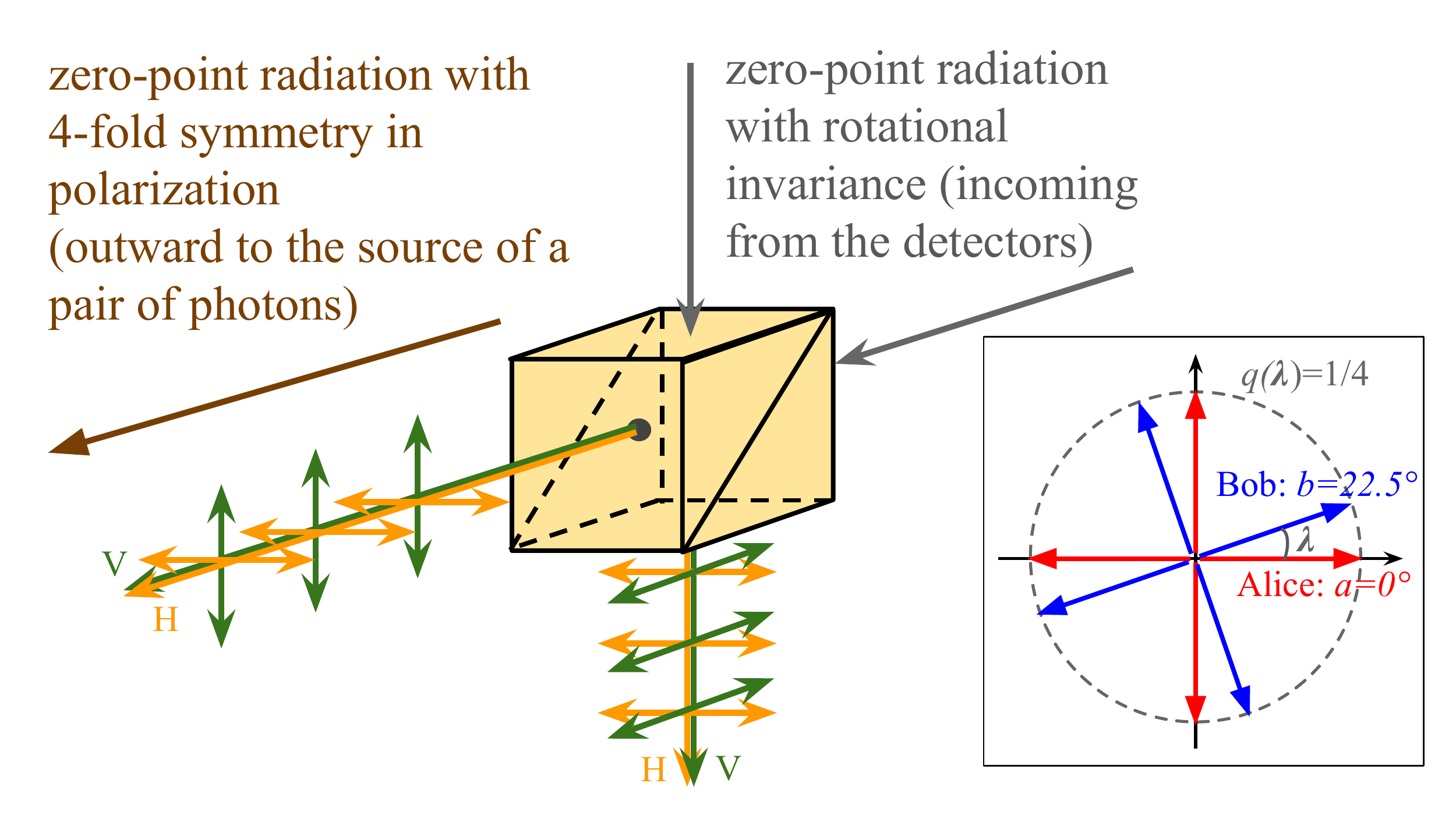}
\caption{A PBS is one of the strongest possible sources of vacuum texture in the typical Bell's inequality measurement. It is very plausible that the PBS would transmit a vacuum texture to the photon source. This vacuum texture does not posses full rotational polarization invariance. Zero-point radiation can be thought of as the cause of vacuum texture. In this model, without a PBS, zero-point radiation caused by vacuum fluctuations will have a uniform distribution of polarization.  However, when the PBS is present, zero-point radiation that impinges on the back of the PBS will become polarized, thus breaking the symmetry. The inset shows $q(\lambda)$ in the polar coordinate in the 4-fold symmetry with the eight delta functions  which are expressed in Eq. \ref{eqn:qab} ($-\pi/2<\lambda\leq\pi/2$). In other words, in terms of the conventional interpretation of quantum field theory, a $\vert 0 \rangle$ photon state is a quantum state just as much as a $\vert 1 \rangle$ photon state; so that, the PBS forms a squeezed vacuum state with large uncertainty in the number of virtual photons in the vacuum. The effects of squeezed $\vert 0 \rangle$ photon states are not negligible especially when the measurement intensity is down to $\vert 1 \rangle$ photon states.} 
\label{fig:PBS}
\end{figure}

\begin{figure}[h]
\centering
\includegraphics[width=0.5\textwidth]{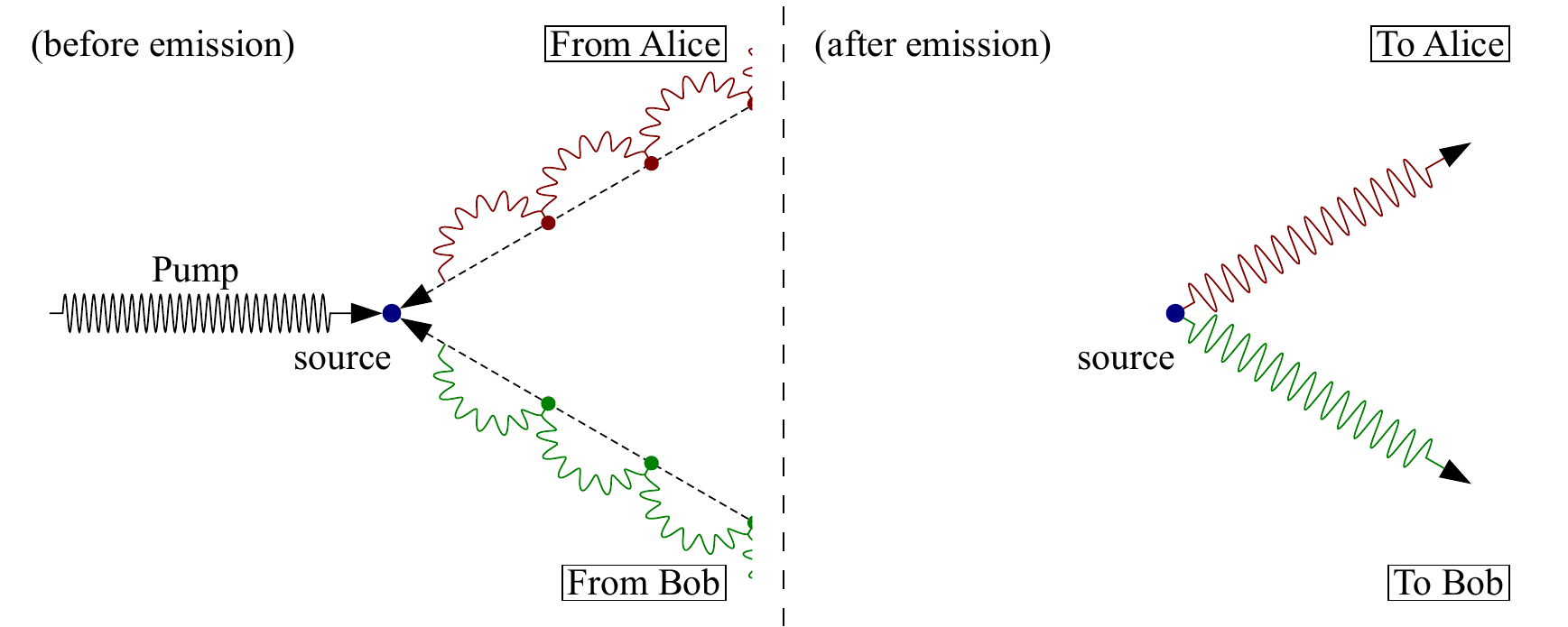}
\caption{A Feynmam diagram depicting a  plausible (but not necessary) model of how a vacuum texture can travel from the PBS to the source: The vacuum texture propagates as a chain of virtual photons. With this model, we require a polarization maintaining interaction between virtual photons. In other words, the presence of one virtual photon with a particular polarization will induce the existence of another virtual photon with the same polarization.  Without the PBSs, the vacuum texture propagating to the source would be unpolarized.  However, the presence of the PBSs only allows certain polarization angles to pass through, creating an inhomogeneous vacuum texture.}
\label{fig:Feynman}
\end{figure}

In Fig. \ref{fig:settingAB} (b), when the pair of photons are created at the source, the polarization of the created photons are no longer rotationally invariant due to the vacuum texture which was created at the PBS and traveled to the source. In Fig. \ref{fig:PBS} and \ref{fig:Feynman}, we propose a plausible model that can lead to a vacuum texture, which for PBS's will produce a vacuum texture with 4-fold symmetry. While any specification of a vacuum texture with 4-fold symmetry is sufficient to determine if it can explain previous Bell's inequality measurement experiments, it is natural to assume that the PBS angles $a$, $a-\pi/2$, $b$ and $b-\pi/2$ degrees are relatively favorable for the first photon to be created. The creation of the secondary photon needs to satisfy conservation of energy and linear \& angular momentum. For simplicity, the delta function has been chosen to describe the local hidden variable with vacuum texture from the PBS with the setting angles of $a$ and $b$:
\begin{multline}
q(\lambda |ab) d\lambda=\frac{1}{4}\{\delta(\lambda-a)+\delta(\lambda-(a-\pi/2))\\
+\delta(\lambda-b)+\delta(\lambda-(b-\pi/2))\}d\lambda,
\label{eqn:qab}
\end{multline}
Using Eq. \eqref{eqn:Eab} and \eqref{eqn:semic}-\eqref{eqn:qab}, the correlation function with vacuum texture is calculated as:
\begin{equation}
E_{vt}(a,b)=\cos(2(a-b)),
\label{eqn:evt}
\end{equation}

Even though we have derived the correlation function assuming a semi-classical model of photons without any entanglement, we arrive at an equation that is identical to quantum mechanical solution that requires entanglement: $E_{vt}(a,b)=E_{qm}(a,b)$.  Thus, we have shown that this deterministic local hidden variable vacuum texture theory agrees with the probabilistic aspect of quantum mechanical theory.

\noindent\fbox{\begin{minipage}{\columnwidth}Thus, tests for the violation of Bell's inequality can't be used to call into question local realism or causality because it is possible to construct a vacuum texture description that yields the exact same quantum mechanical inequality.  
\end{minipage}}

\section{\label{sec:FreedomOfChoice}Freedom of Choice}
\begin{figure}[h]
\centering
\includegraphics[width=0.45\textwidth]{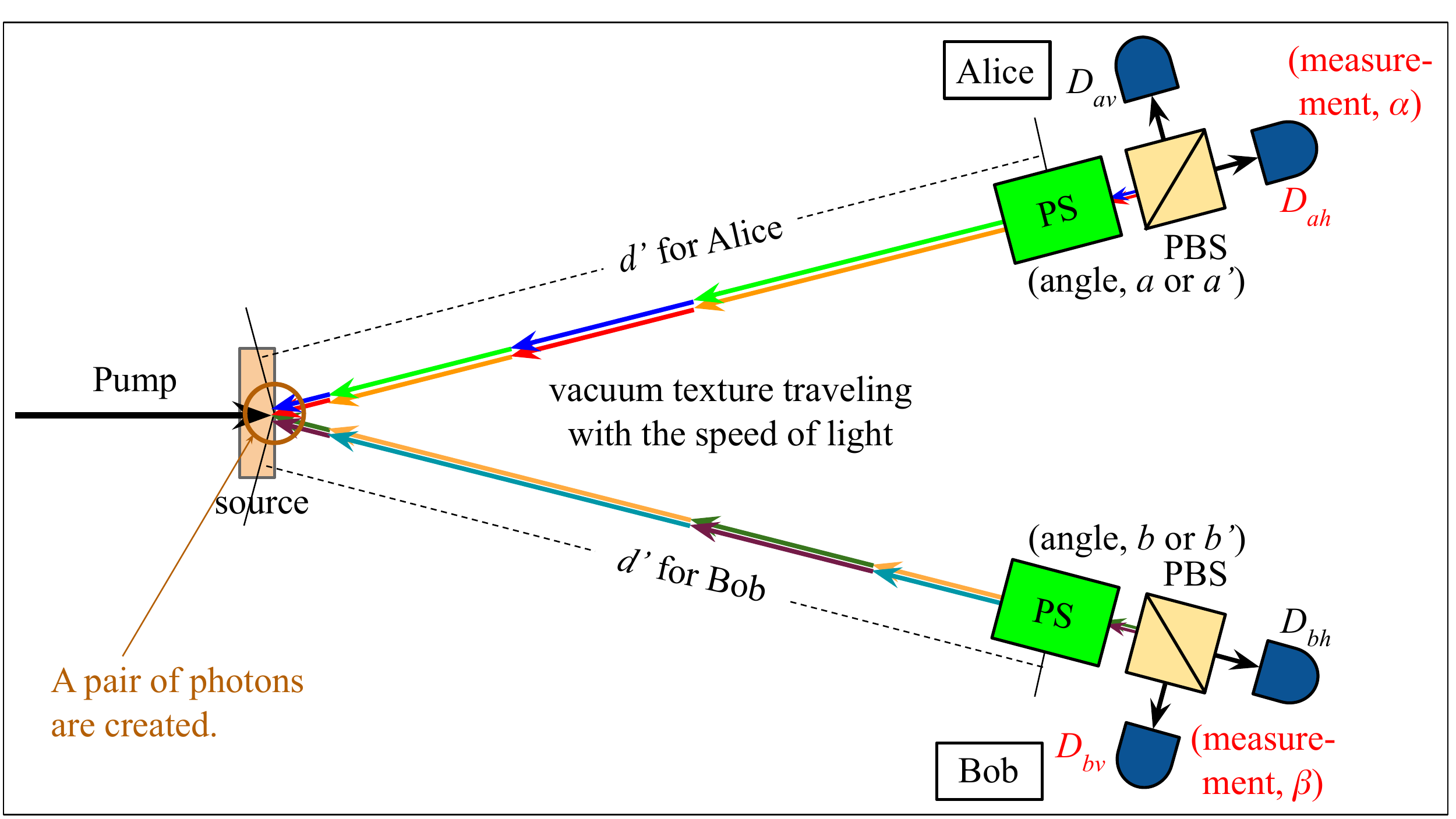}
\caption{Many different schemes have been used to measure Bell's inequality.  Here we show a simplified scheme involving a phase shifter (PS) used as a switch for the measurement settings of polarization angles between $a$ \& $a'$ for Alice and $b$ \& $b'$ for Bob. However, it is important to understand that regardless of the experimental details there will always exist a vacuum texture traveling from the measurement device back to the source which influences the production of the photons being measured. Also, unless otherwise stated, it is assumed that $d'\approx d$ in Fig. \ref{fig:settingAB} (b). }
\label{fig:settingC}
\end{figure}

Bell was highly aware of the possibility that $q(\lambda |a,b)\neq q(\lambda )$ \cite{Bell, Bell1987}. Consequently a series measurements were performed that added the feature of freedom of choice \cite{AspectDec, Tittel, Weihs, Giustina, Shalm}. Two choices/settings are made after/before the creation of a pair of photons at the source either periodically or randomly. Alice had a choice between $a$ or $a'$, and Bob had a choice between $b$ or $b'$ (see Fig. \ref{fig:settingC}). In the vacuum texture theory with freedom of choice, the following sequence of events occurs: 
\begin{enumerate}
  \item Alice/Bob choose a detector setting
  \item The vacuum texture travels from the detector to the source in time,  $d/v_{vt}\approx d/c$
  \item The vacuum texture interacts with the source for an interaction time, $\tau$ (we assume $\tau \ll d/c$).
  \item The photons (with polarization influenced by the vacuum texture/detector setting) travels to each of the detectors in time, $d/c$.
\end{enumerate}

The probability function within the vacuum texture theory with choice becomes:
\begin{multline}
P_{vt}(\alpha \beta|a_m b_m\,(a_v b_v))\\
=\int^{\pi/2}_{-\pi/2}d\lambda q(\lambda |a_vb_v)P(\alpha|a_m,\lambda )P(\beta|b_m,\lambda)
\label{eqn:Pvtabab}
\end{multline}
where $a_m$ (which can either be $a$ or $a'$) and $b_m$ (which can be either $b$ or $b'$) are the measurement settings which were actually used for the measurement results of $\alpha$ and $\beta$, and $a_v$ (which can be either $a$ or $a'$) and $b_v$ (which can be either $b$ or $b'$) are the measurement settings when they are in the past lightlike cone of the pair-creation event.  In other words,  $P_{vt}(\alpha \beta|a_m b_m\,(a_v b_v))$ the probability of measuring $\alpha$ and $\beta$ given the PBS's were in state $a_v$ and $b_v$ at the moment the vacuum texture left the PBS's \textit{and} the PBS's were in state $a_m$ and $b_m$ at the moment the photon pair emitted from the sample arrive at the PBS's.

Because of the round-trip time, there are two possible measurement scenarios when $a$ \& $a'$ and $b$ \& $b'$ are changing: in-sync and out-of-sync.  The in-sync scenario for Alice occurs when $a_m=a_v$ and out-of-sync scenario occurs when $a_m\neq a_v$, with similar conditions for Bob. In experiments conducted to date, Alice and Bob switch their detector between two different values either periodically or randomly. In either case, the probabilities are accumulated over a sufficiently long time interval so that there will be a fraction of time, $f_A$, when Alice has $a_m=a_v$, in-sync scenario and $1-f_A$ when she has $a_m\neq a_v$, out-of-sync (with similar definitions for Bob). Here, we assume that $f_A$ is same for $a_m=a$ and $a_m=a'$ ($f_B$ is same for $b_m=b$ and $b_m=b'$). This case is especially important for the case for periodic switching which is discussed in the next section. 

Under these conditions, the distribution of the hidden variable given that the measurement setting are $a$, $b$ and in the past they could have been $a$, $a'$, $b$, $b'$ has the following form:
\begin{multline}
q^{fc}(\lambda|ab\,(aba'b'))d \lambda=
\{f_Af_B\,q(\lambda |a b)+f_A(1-f_B)q(\lambda |ab')\\
+(1-f_A)f_B\,q(\lambda |a' b)+(1-f_A)(1-f_B)q(\lambda |a' b')\} d \lambda\\
=\{fq(\lambda |ab)+(1-f)q(\lambda |a'b')+f'(q(\lambda |ab')-q(\lambda |a'b))\} d \lambda.
\label{eqn:qabab}
\end{multline}
Where in the last line we used Eq. \ref{eqn:qab}, and have $f=(f_A+f_B)/2$ and $f'=(f_A-f_B)/2$.\footnote{With vacuum texture in the case of periodic choice, when $\nu_A=n\nu_B$, n=1,2,3... (or $\nu_B=n\nu_A$, n=1,2,3...) where $\nu_A$ and $\nu_B$ are frequencies of periodic switching for Alice and Bob, $f_A$ and $f_B$ are no longer independent of each other. In stead, they are correlated and depend on their relative phase. In Eq. \ref{eqn:qabab}, we assumed that $f_A$ and $f_B$ are independent. However, using Eq. \ref{eqn:qab}, the last line of Eq. \ref{eqn:qabab} still holds for the case of periodic choice regardless of the relative phase, and the equations are same as the case of freedom-of-choice.} With Eq. \ref{eqn:Pvtabab}, we find that the probability for freedom-of-choice when the measurement settings are $a$ and $b$ under vacuum texture theory becomes 
\begin{multline}
P_{vt}^{fc}(\alpha \beta|ab\,(aba'b'))\\
=\int^{\pi/2}_{-\pi/2}d\lambda q^{fc}(\lambda |ab\,(aba'b'))P(\alpha|a,\lambda )P(\beta|b,\lambda)\\
=fP_{vt}(\alpha \beta|ab\,(ab))+(1-f)P_{vt}(\alpha \beta|ab\,(a'b'))\\
+f'\{P_{vt}(\alpha \beta|ab\,(ab'))-P_{vt}(\alpha \beta|ab\,(a'b))\}.
\label{eqn:Prcvt}
\end{multline}

Consequently, the correlation value under freedom-of-choice for actual measurement settings $a$ and $b$ in an experiment that has the possible free choices $a$, $b$, $a'$, $b'$ is expressed as:
\begin{multline}
E^{fc}_{vt}(a,b\,;a',b)=\sum_{\alpha,\beta=\pm 1}\alpha \beta P^{fc}_{vt}(\alpha \beta|ab\,(aba'b'))=\\
f E^{is}_{vt}(a,b)+(1-f)E^{os}_{vt}(a,b\,;a',b')+f'E^{ub}_{vt}(a,b\,;a',b').
\label{eqn:Efc}
\end{multline}
where $E^{is}_{vt}(a,b)$ represents the contribution when Alice and Bob's detectors are both in sync. $E^{os}_{vt}(a,b\,;a',b')$ represents the contribution when when both their detectors are out of sync. The third term could be non-zero only when the system is unbalanced: $f_a\neq f_b$ ($f'\ne0$).

Within vacuum texture theory, for the in-sync term, we revert to the standard correlation value:
\begin{equation}
E^{is}_{vt}(a,b)=\sum_{\alpha,\beta=\pm 1}\alpha \beta P_{vt}(\alpha \beta|ab\,(ab))=\cos(2(a-b)).
\end{equation}
For the out-of-sync term: \begin{multline}
E^{os}_{vt}(a,b\,;a',b')=\sum_{\alpha,\beta=\pm 1}\alpha \beta P_{vt}(\alpha \beta|ab\,(a'b'))\\
=\frac{1}{2}\{\cos(2(a-b))+\cos(2(a+b-a'-b'))\cos(2(a'-b'))\}.
\label{eqn:Eos}
\end{multline}

For the unbalanced term,
\begin{multline}
E^{ub}_{vt}(a,b\,;a',b')\\=\sum_{\alpha,\beta=\pm 1}\alpha \beta \{P_{vt}(\alpha \beta|ab\,(ab'))-P_{vt}(\alpha \beta|ab\,(a'b))\}\\
=\frac{1}{2}\{\sin(2(a'+b'-a-b))\sin(2(a'-b'))\}.
\label{eqn:Eub}
\end{multline}

The experiments often measure Bell's $S$ value for $a=0$, $b=\pi/8$, $a'=\pi/4$ and $b'=3\pi/8$ where the Bell-CHSH inequality \cite{CHSH} is expressed as $|S|\leq2$. It is easy to show that:
\begin{equation}
S^{fc}_{vt}(0,\pi/8,\pi/4,3\pi/8) = 2\sqrt{2} f ,
\label{equ:Sfc}
\end{equation}
where
\begin{multline}
S^{fc}_{vt}(a,b,a',b')=|E^{fc}_{vt}(a,b\,;a',b')-E^{fc}_{vt}(a,b'\,;a',b)\\
+E^{fc}_{vt}(a',b\,;a,b')+E^{fc}_{vt}(a',b'\,;a,b)|.
\end{multline}
From the above analysis, it is clear that depending upon the experimental details of the particular freedom-of-choice experiment, under the vacuum texture theory, Bell's $S$ value can range anywhere from the quantum value of $2\sqrt{2}$, to the semi-classical value of $\sqrt{2}$, and all the way to $0$.

If the choice is completely random (like  most of the experiments were intended \cite{Tittel, Weihs, Giustina, Shalm}), $f=1/2$ and $f'=0$ so that Eq. \ref{eqn:Efc} simplifies to:
\begin{multline}
E^{rc}_{vt}(a,b\,;a',b')=
\frac{E^{is}_{vt}(a,b)+E^{os}_{vt}(a,b\,;a',b')}{2}.
\label{eqn:Erc}
\end{multline}
The Bell's S value for random choice :
\begin{equation}
S^{rc}_{vt}(0,\pi/8,\pi/4,3\pi/8)=\sqrt[]{2} \approx 1.41 < 2.
\end{equation}
Interestingly, this equals the value expected from a semi-classical model, $S_{sc}=\sqrt{2}$.  This is not consistent with many freedom-of-choice measurements \cite{Tittel, Weihs, Giustina, Shalm}. However, those measurements, involving single-photon intensities extended to many kilometers, suffer from very poor sampling rates. Which in turn decreases our ability to make any definitive claims.  For example, a fair sampling assumption is applied to the data. Shown in Fig. \ref{fig:Angle}, the semi-classical result (blue curve) which can be made to resemble the quantum one (red curve) simply by scaling the data as is needed to impose fair sampling. 

Moreover, in the vacuum texture model, as the signal-to-noise ratio decreases, the signals depend more on the vacuum texture itself which is created by the various experimental apparatus near the detector and depends less on the consequence of the incident excitation (i.e. the action of the pump beam on the source) as many low-signal experiments depend on timed-window detection rather than signal-triggered detection. In any cases, random choice is not the best measurement setting to prove or disprove the existence of the vacuum texture.  A more suitable measurement is suggested in the next section.

\section{\label{sec:Experiment}Suggested Experiment}
In 1982, Aspect et al. performed the most cited Bell's inequality measurement \cite{AspectDec}. This was the first experiment to successfully introduce \textit{choice}. The PBS angles of $a$ and $a'$ for Alice, and $b$ and $b'$ for Bob were switched periodically at a frequency around 50 MHz by introducing a phase delay in each path via an electro-optical phase modulator \cite{Aspect2002, Aspect2004}.

In the 1982 Aspect experiment, a different form of Bell's inequality, $S'$ was used:
\begin{multline}
S'=\frac{N(a,b)}{N(\infty,\infty)}-\frac{N(a,b')}{N(\infty, \infty')}+\frac{N(a',b)}{N(\infty', \infty)}+\frac{N(a',b')}{N(\infty', \infty')}\\
-\frac{N(a',\infty)}{N(\infty', \infty)}-\frac{N(\infty,b)}{N(\infty, \infty)},
\label{eqn:sprime}
\end{multline}
where $N(\infty,\infty)$ is without a PBS, and $N(a,b)/N(\infty, \infty) \equiv n(a,b)=P(\alpha=\beta=1|ab)$ is the probability of a photon being detected with a single detector measurement for both Alice and Bob with the PBS setting angle $a$ and $b$ and with Malus's law, $P(\alpha=1|a,\lambda )d\lambda=\cos^2(a-\lambda)d\lambda$. Therefore, in theory, $N(a,\infty)/N(\infty, \infty)=N(\infty,a)/N(\infty, \infty)=1/2$. The inequality for a local hidden variable in Eq. \ref{eqn:lhv} is expressed as $-1\leqslant S'\leqslant 0$.

\begin{figure}[h]
\centering
\includegraphics[width=0.45\textwidth]{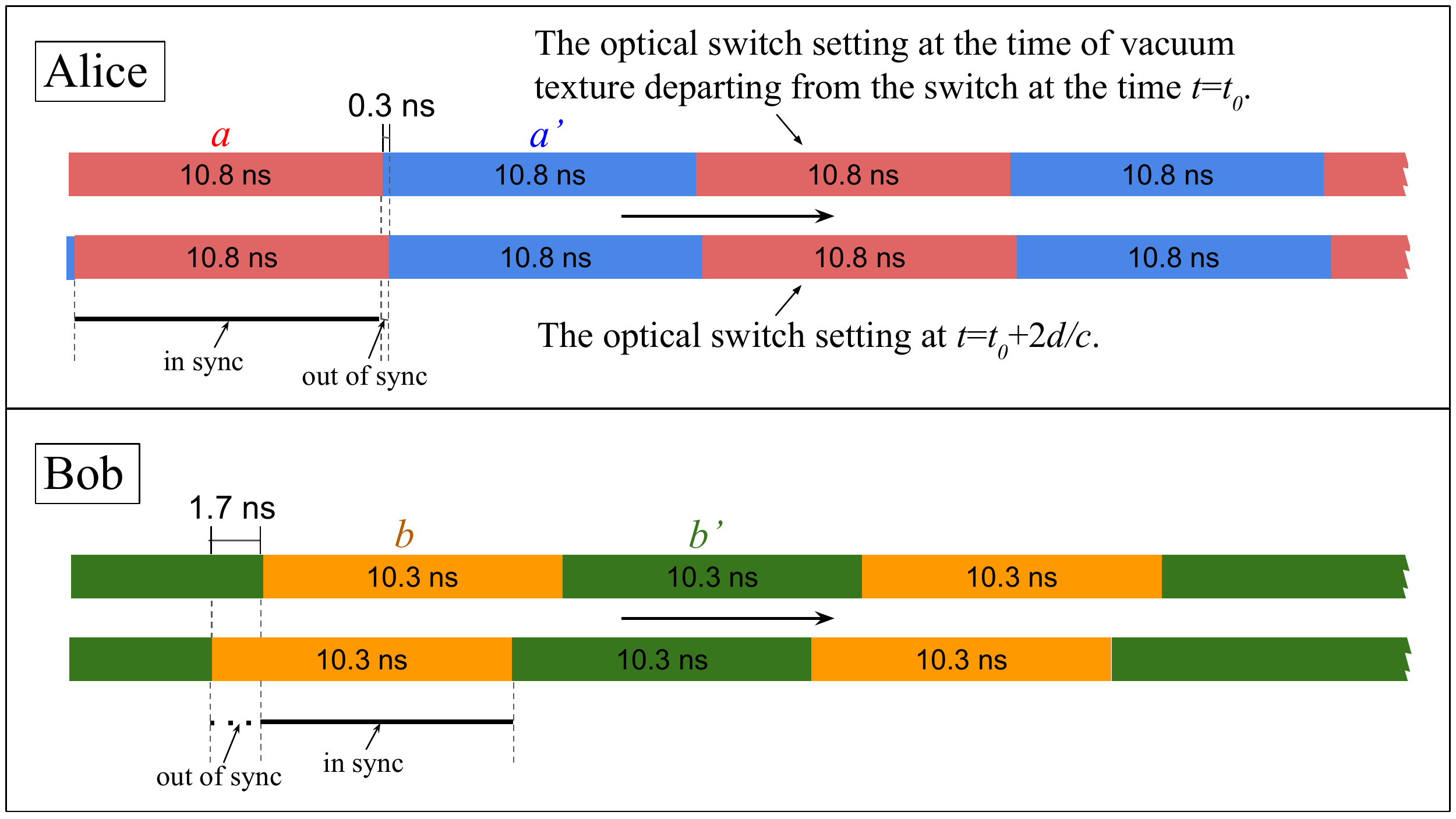}
\caption{In the Aspect experiment, $f_A=0.97$, $f_B=0.83$ ($f=0.90$ and $f'=0.07$).}
\label{fig:Choice}
\end{figure}

With $a$, $b$, $a'$ and $b'$ of $0$, $\pi/8$, $\pi/4$ and $3\pi/8$, respectively, Eq. \ref{eqn:sprime} simplifies to $S'_{qm}=0.207$ for the standard quantum mechanical interpretation, since $n_{qm}(a,b)=\cos^2(a-b)/2$. With vacuum texture without choice we also have, $S'_{vt}=0.207$ since $n_{vt}(a,b)=\cos^2(a-b)/2$. For the semi-classical case, $S'_{sc}=-0.146$ where $n_{sc}(a,b)=\frac{1}{8}(2+\cos(2(a-b)))$. 

Similar to Eq. \ref{eqn:Efc}, under vacuum texture with freedom-of-choice, we have
\begin{multline}
n^{fc}_{vt}(a,b\,;a',b')=f n^{is}_{vt}(a,b)+(1-f)n^{os}_{vt}(a,b\,;a',b')\\
+f'n^{ub}_{vt}(a,b\,;a',b')
\label{eqn:nfc}
\end{multline}
where $n^{is}_{vt}(a,b)=n_{qm}(a,b)=\cos^2(a-b)/2$ is the in-sync component, $n^{os}_{vt}(a,b\,;a',b')=P_{vt}(\alpha=\beta=1|ab\,(a'b'))=\frac{1}{8}\{2+\cos(2(a-b))+\cos(2(a+b-a'-b'))\cos(2(a'-b'))\}$ is the out-of-sync component, and $n^{ub}_{vt}(a,b\,;a',b')=P_{vt}(\alpha= \beta=1|ab\,(ab'))-P_{vt}(\alpha= \beta=1|ab\,(a'b))=\frac{1}{8}\{\sin(2(a'+b'-a-b))\sin(2(a'-b'))\}$ is the unbalanced component.

Therefore, with $a$, $b$, $a'$ and $b'$ of $0$, $\pi/8$, $\pi/4$ and $3\pi/8$,,
\begin{equation}
S'^{fc}_{vt}(0,\pi/8,\pi/4,3\pi/8)=-\frac{1}{2}+\frac{f}{\sqrt{2}}
\label{equ:Sps}
\end{equation}
where
\begin{multline}
S'^{fc}_{vt}(a,b,a',b')=n^{fc}_{vt}(a,b\,(a',b'))-n^{fc}_{vt}(a,b'\,(a',b))\\
+n^{fc}_{vt}(a',b\,(a,b'))+n^{fc}_{vt}(a',b'\,(a,b))-\frac{1}{2}-\frac{1}{2}.
\end{multline}

For the periodic switching case, we can obtain $f_A$ (and $f_B$) from the switching frequencies $\nu_A$ (and $\nu_B$). For Alice, $f_A$ has the maximum condition if an integral multiple of the wave length would fit in the light path, $2d$ which is the round trip distance between PBS and the source. 
\begin{equation}
\begin{split}
2d=\lambda_A n \rightarrow f_A=1\\
2d=\lambda_A (n+\frac{1}{2}) \rightarrow f_A=0
\end{split}
\label{eqn:n}
\end{equation}
where $\lambda_A=c/\nu_A$ and $n=0,1,2,3...$ If we assume the switching to be a perfect square wave, for Alice,
\begin{equation}
f_A=\frac{1}{\pi}\arccos{(\cos{\{2\pi(\frac{2d}{c}\nu_A-\frac{1}{2})\}})},
\label{equ:tw}
\end{equation}
with similar definitions for Bob.

\begin{figure}[h]
\centering
\includegraphics[width=0.5\textwidth]{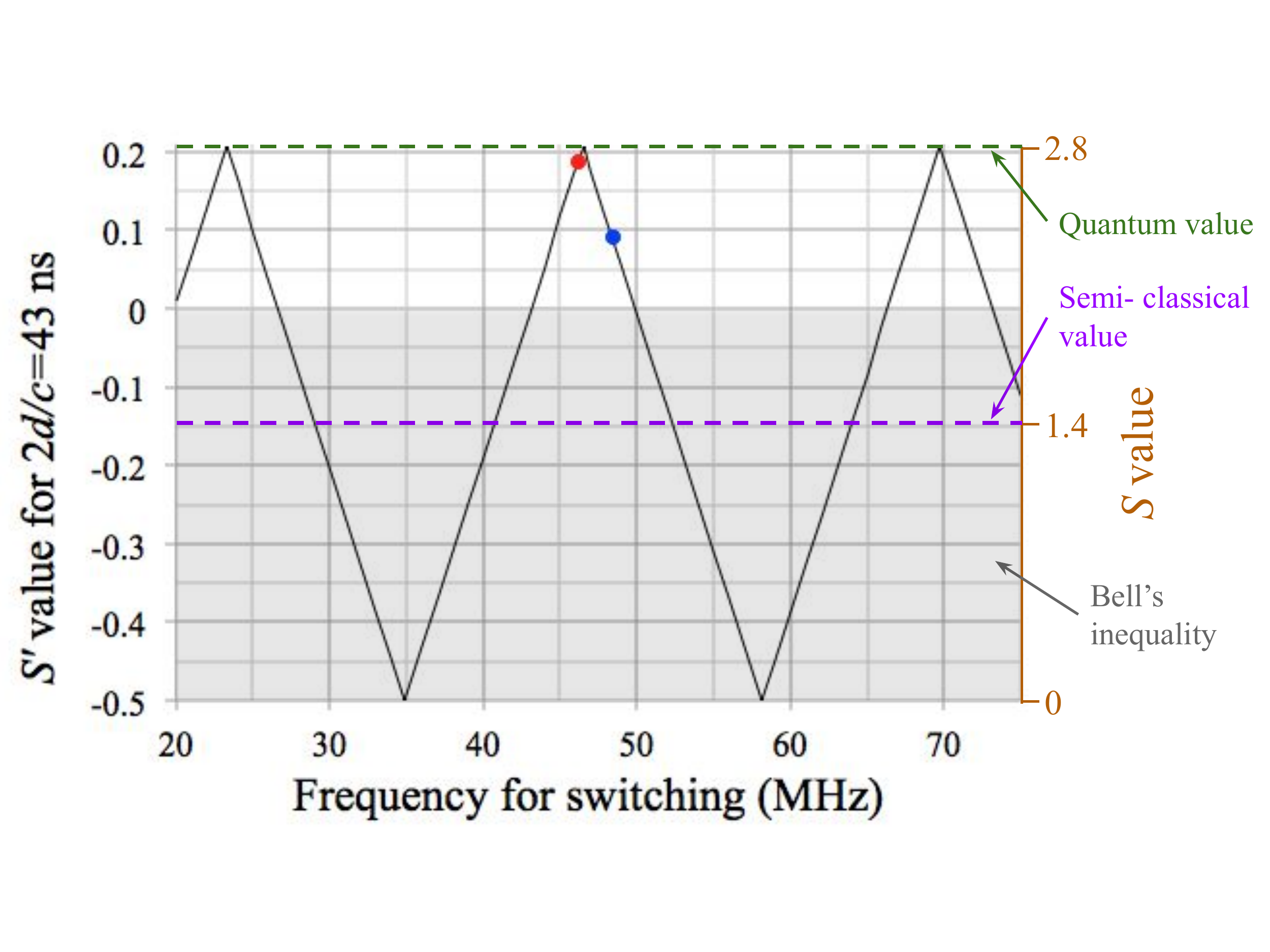}
\caption{The expected $S'$ value with vacuum texture is plotted in a function of switching frequency. The red dot is at the frequency which was used for Alice, and the blue dot is at the frequency which was used for Bob in \cite{AspectDec}. The measured $S'$ is the average of the settings of Alice and Bob (see Eq. \ref{equ:Sps} and \ref{equ:tw}). Expected $S$ value in Eq. \ref{equ:Sfc} under the same conditions is also shown in orange. The green dotted line represents the quantum value. The purple dotted line represents the semi-classical value. The shaded region shows Bell's inequality.}
\label{fig:FreqSweep}
\end{figure}

In  the  1982  Aspect  experiment, the round trip time for light between the switch and the pair of photon source was $2d/c=43$ ns for Alice and Bob. The optical switching frequency for Alice was 46.2 MHz and then the switching period was 43.3 ns. The optical switching frequency for Bob was 48.4 MHz and then the switching period was 41.3 ns \cite{AspectDec,Aspect2002, Aspect2004}. In 1989, Zeilinger pointed that ``there was a numerical coincidence between photon flight time and switching frequency \cite{Zeilinger1986}.'' In this measurement, the average of Alice and Bob had 90 \% of the time in phase, $f=0.9$ (see Fig. \ref{fig:Choice}). Therefore, the predicted value with vacuum texture is $S'^{fc}_{vt}=0.136$. Any experimental error in the switching in the experiment would suppress the value of $S'$ by a small amount. The measured value in ref. \cite{AspectDec} was $S'_{measured}=0.101\pm0.020$, in reasonable agreement with the theory of vacuum texture.

In order to prove or disprove the existence of vacuum texture. We suggest to repeat the same or a similar measurement as conducted by Aspect. However this time, to vary the switching frequency (and/or the distance) to see whether the in-sync and out-of-sync follow the maxima and minima in Fig. \ref{fig:FreqSweep} predicted by the vacuum texture theory. The maxima up to $S'=0.207$ ($S=2.83$) are quantum effects in which, in this new interpretation, the incident particles are in-sync with vacuum texture. At the minima down to $S'=-0.5$ ($S=0$), which corresponds to the completely out-of-sync situation, even classical correlation is diminished. If the frequency setting are random and averaged, the in-sync and out-of-sync components would average out to the semi-classical behavior, $S'_{sc}=-0.146$ ($S=1.41$). 

Furthermore, in this article, so far we have assumed that $d'$ (see Fig. \ref{fig:settingC}) or $d$ without a phase shifter (see Fig. \ref{fig:settingAB} (b)) for Alice and Bob are equal. When they are different, we might be able to observe the dependence of the effects of the vacuum texture on the distance. For example, the switch for Alice can be removed, and Alice's setting can be fixed at $a$ or $a'$ with distance $d_a$ for a set of measurements while Bob can have the same periodical switching between $b$ and $b'$ with distance $d'_b$. At $d_a\gg d'_b$, the frequency sweep for Bob should show the maxima and minima with the same periodicity in Fig. \ref{fig:FreqSweep} in $S'$. Under these circumstances the delta functions appearing in Eq. \ref{eqn:qab} will have unequal weights for Alice and Bob.  The maxima and minima in $S'$ should disappear in the limit where $d_a\ll d'_b$, and the value should be constant. Measurements taken when $d_a\approx d'_b$ would be very interesting because it would enable us to determine the amplitudes of the weight functions as a function of distance appearing in Eq. \ref{eqn:qab}.

\section{\label{sec:discussion}discussion}

In our interpretation, we have shown that  quantum nonlocality such as superposition and entanglement are not needed to describe Bell's inequality measurements and perhaps other measurements as well. Vacuum texture (a local variable) may also easily explain away the so-called observations of quantum-to-classical transition such as dephasing. As we showed, the interaction with vacuum texture (a local variable) is in-sync or out-of-sync and results in maxima or minima of expectation values of measured quantities. If the settings of an experimental apparatus, during the experiment, is unstable or in an uncontrollable environment due to effects such as thermal fluctuations, the effects of the vacuum texture (a local variable) on the expectation values will become washed out. Consequently, the in-sync and out-of-sync effects are averaged out to the classical value within the dephasing time, $T_2$ . In the classical world, the settings of the environment would change due to thermal fluctuation as the vacuum texture travels between masses. This could be the reason that quantum effects are often observed only at low temperatures in relatively small systems.

In Fig. \ref{fig:FreqSweep}, there is no frequency dependence in the conventional interpretation of quantum mechanise (green dotted line). Our vacuum texture interpretation with a local variable with complete causality reveals more detailed information (black solid lines) with maxima and minima. The semi-classical value is at the averaged value when the maxima and minima structure is washed out. The most cited Bell's inequality measurement of 1982 Aspect's measurement \cite{AspectDec} has never been reproduced by anyone. The non-constant S values at different frequencies might startle experimentalists. However, this could be the most important information in physics over the past century.

\section{\label{sec:concludion}Conclusion}
We have proposed that the existence of vacuum texture would explain Bell's inequality measurements with local realism and causality. An experiment, which was capable with 1982 technology, is suggested for definite proof/disproof of vacuum texture. If the existence of the vacuum texture is proven by experiment, we will finally have the definitive  answer to the EPR question, and a local variable theory should be back on the table as a possible interpretation of quantum mechanics.

\begin{acknowledgments}
This work was supported by the M. Hildred Blewett Fellowship of the American Physical Society, www.aps.org and the Laboratory Directed Research and Development program of Los Alamos National Laboratory under project No. 20160584ER.  We would like to thank Steve Buelow, Myriam Sarachik and Eugene Chudnovsky.
\end{acknowledgments}

\end{document}